\documentclass{IEEEtran}

\usepackage{setspace}
\usepackage{mathptmx}
\usepackage{amsmath}
\usepackage{amssymb}
\usepackage{epsfig}
\usepackage{amsthm}
\usepackage{amsfonts}
\usepackage{subfigure}
\usepackage{stmaryrd}
\usepackage[dvips]{color}
\usepackage{graphicx}
\usepackage{mathcomSTEP}

\allowdisplaybreaks

\newtheorem{definition}{Definition}

\theoremstyle{remark}
\newtheorem{rem}[thm]{Remark}

\begin{document}

\title{
The Impact of Phase Fading \\
on the Dirty Paper Channel
}

\author{%
\authorblockN{%
Stefano~Rini\authorrefmark{1} and Shlomo~Shamai~(Shitz)\authorrefmark{2} \\
}

\authorblockA{%
\authorrefmark{1}
Stanford University, Stanford, CA, USA \\
E-mail: \texttt{stefano@wsl.stanford.edu} }

\authorblockA{%
\authorrefmark{2}
Technion-Israel Institute of Technology,  Haifa, Israel \\
E-mail: \texttt{sshlomo@ee.technion.ac.il} }

\thanks{
The work of S. Rini was partially funded by the  NSF Center for Science of Information (CSoI) under grant CCF-0939370.
The work of S. Shamai was supported by the Israel Science Foundation (ISF) and by the European FP7 NEWCOM
%
}
}

\maketitle

\begin{abstract}
The impact of phase fading on the classical Costa's dirty paper coding channel is studied.
We consider a variation of this channel model in which the amplitude of the interference sequence is known at the transmitter while
its phase is known at the receiver.
Although the capacity of this channel has already been established, it is expressed using an auxiliary random variable and as the solution of a maximization problem.
%
To circumvent the difficulty evaluating capacity, we derive alternative inner and outer bounds and show that the two expressions are to within a finite distance.
This provide an approximate characterization of the capacity which depends only on the channel parameters.
%
%
We consider, in particular, two distributions of the phase fading: circular binomial and circular uniform.
%
The first distribution models the scenario in which the transmitter has a minimal uncertainty over the phase of the interference while the second distribution models complete uncertainty.
For circular binomial fading, we show that binning with Gaussian signaling still approaches capacity, as in the channel without phase fading.
In the case of circular uniform fading, instead, binning with Gaussian signaling is no longer effective and novel interference avoidance strategies are developed to approach capacity.
%
%
%
%
\end{abstract}
%
%

\section{Introduction}
%
\IEEEPARstart{W}{ith}
 the increase in network traffic and density, Base Station (BS) cooperation is becoming a common feature of  modern cellular communication system.
BS cooperation offers many advantages: for instance, coordinated multi-point transmissions provide crucial coherent combining gains for users on the cell edge.
Another advantage provided by BS cooperation is interference pre-cancellation: having knowledge of the interference created by neighbouring BSs at the intended receiver, a BS can pre-code its transmission against such interference.
The information theoretic model which characterizes the limiting performance of interference pre-cancellation is the Gel'fand-Pinsker (GP) problem \cite{GelFandPinskerClassic}.
%
%
Although well understood in the literature, the GP problem is rarely considered in practical systems.
The difficulty in translating this theoretical results into practical transmission strategies partially lies in the idealized assumption that the transmitter has perfect knowledge of the communication channel.
%
%
%
Channel knowledge at the BS is particular hard to obtain for different reasons, the main of which is perhaps fading.
%
%
In this paper, we address the effect of partial transmitter channel knowledge in the presence of phase fading and characterize the optimal transmission strategies for different distributions of the fading realizations.

In the GP channel, a transmitter communicates to a receiver over a channel subject to both noise and state: the state is known non-causally at the transmitter but is not known at the receiver.
%
%
A variation of the classical GP problem \cite{GelFandPinskerClassic} is the model in which the channel state is partially known at the transmitter and partially known at the receiver.
The capacity of this more general channel is established by Cover and Chiang in \cite{cover2002duality}.
%
The GP problem in which the channel output is obtained as a linear combination of the input, the state which models interference and a white Gaussian noise is considered by Costa \cite{costa1983writing}.  For this channel, it is shown that the presence of the interference does not reduce capacity: this celebrated result is known as  ``writing on dirty paper''.
The variation of the writing on dirty paper channel in which fading is added to the interference sequence is known as  ``writing on fading dirt''.
The capacity of this channel is a special case of \cite{cover2002duality}  but its expression contains an auxiliary random variable and is obtained as the solution of a maximization problem.
For these reasons, neither closed form expressions nor numerical evaluations of the capacity for the writing on fading dirt problem are not known.
Outer and inner bounds to the capacity of the writing on fading dirt channel are derived in \cite{grover2007need,grover2007writing} while achievable rates under Gaussian signaling and lattice strategies are derived in \cite{avner2010dirty}.
An outer bound for the vector writing on fading dirt problem was recently derived in \cite{kao2013upper}.
%
%
%

In the following, we focus on the writing on fading dirt problem for the case in which only phase fading is considered.
Additionally, the phase fading process is assumed to be known at the receiver but not at the transmitter.
%
%
We study the capacity of this channel for two distributions of the phase fading: the circular binomial and circular uniform distribution.
The first distribution represents the case in which the uncertainty over the fading process is minimal while the second distribution the case in which it is maximal.
In both cases, we derive new inner and outer bounds and show that they lie to within finite additive gap which does not depend on the channel parameters.
%
%
For the binomial circular distribution, the scheme which approaches capacity relies on binning with Gaussian signaling as in the channel without fading.
%
For the circular uniform distribution, a novel transmission strategy is developed in which the transmitter only uses one dimension to send information while the other dimension is used to estimate the interference.
The rest of the paper is organized as follows.
Section \ref{sec:Dirty Paper Channel with Phase Fading} introduces the channel model. Section \ref{sec:Related Results} present the relevant results derived in the literature.
In Section \ref{sec:Two Phase Fading Values}, we study the circular binomial phase fading case while, in Section  \ref{sec:circular uniform Phase Fading}, we investigate the circular uniform case.
Section \ref{sec:Numerical Simulations} presents relevant numerical simulation. Finally, Section \ref{sec:Conclusion} concludes the paper

In the following only sketches of the proofs appear: full proofs can be found in an extended version available online \cite{RiniPhase14}.

\section{Dirty Paper Channel with Phase Fading}
\label{sec:Dirty Paper Channel with Phase Fading}
\begin{figure}
\centering
\includegraphics[width=.54 \textwidth]{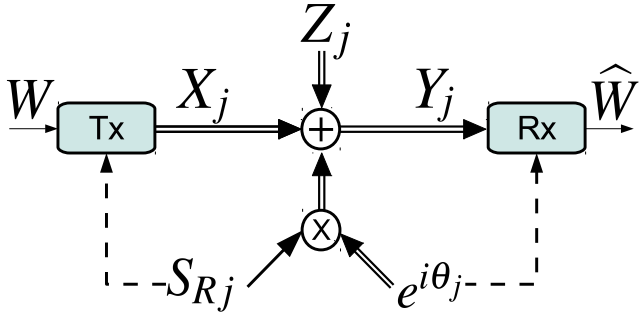}
\vspace{-3.9 cm}
\caption{The Dirty Paper Channel with Phase Fading (DPC-PF) model.
The single line indicates real values while the double line indicates complex values.
}
\vspace{-.5 cm}
\label{fig:WritingOnFadingDirt}
\end{figure}
In Dirty Paper Channel with Phase Fading (DPC-PF), also depicted in Fig. \ref{fig:WritingOnFadingDirt}, the channel output is obtained as
\ea{
Y_j = X_j + e^{i \theta_j} S_{Rj} + Z_j,
\label{eq:WritingOnFadingDirtChannelModelGeneral}
}
for  $i=\sqrt{-1}$, $j \in [1 \ldots N]$ and where $X_j$ is the channel input, $S_j$ the interference,  $Z_j$ the additive noise and $\theta_j$ the fading realization.
The channel input $X_j=X_{Rj}+i X_{Ij}$ is subject to the power constraint
\ea{
\Ebb \lsb |X_j|^2 \rsb  =  \Ebb \lsb X_{Rj}^2 + X_{Ij}^2 \rsb  \leq P,
}
and the interference $S_{Rj}$ is a normal Random Variable (RV) with zero mean and covariance $Q$, also indicated as $\Ncal(0,Q)$.
%
%
The noise term $Z_j=Z_{Rj}+i Z_{Ij}$ is an iid circular symmetric complex normal RV with zero mean and unitary covariance, also indicated as $\Ccal\Ncal(0,1)$.
The interference sequence $S_R^N$ is assumed to be anti-causally available at the transmitter while the phase fading sequence $\theta^N$ is known at the receiver.
The term $\theta_j$ represents the effect of phase fading on the interference sequence $S_R$ and is iid draw from the circular distribution $P_\theta$.
%
%
%
In the following we focus on two distributions for $P_{\theta}$:

\begin{itemize}
\item  a circular binomial distribution
\ea{
P_{\theta}(t)=\f 12 \lb 1_{\{t=+\Delta\}}(t) + 1_{\{t=-\Delta\}}(t)\rb, \quad \Delta \in [0, \pi/2]
\label{eq: distribution 2 values}
}
where $1_{\{x  \ \in \ I\}}(x)$ be the indicator function for the set $I$.
\item  a circular uniform distribution
\ea{
P_{\theta}(t)=\f 1 {2\pi}, \quad t \in [0,2 \pi)
\label{eq: distirbution uniform}
}
\end{itemize}
Note that any channel with a circular binomial phase fading can be reduced to the distribution in \eqref{eq: distribution 2 values}
without loss of generality by pre-rotating the channel input and rotating the channel output.
The fading model in \eqref{eq:WritingOnFadingDirtChannelModelGeneral} is usually referred to as ergodic fading or fast fading, since the fading realization changes at each channel use in a memoryless fashion.
This model represents a worst-case scenario: models is which the fading process has memory over the channel uses and vary with less randomness can be obtained from \eqref{eq:WritingOnFadingDirtChannelModelGeneral} by providing the transmitter with a genie-aided side information on the fading process.

\section{Related Results}
\label{sec:Related Results}

The capacity for the DPC-PF in \eqref{eq:WritingOnFadingDirtChannelModelGeneral}  is a special case of the result in  \cite{cover2002duality}.

\begin{thm}{\bf Capacity of the DPC-PF \cite[Th. 1]{cover2002duality} \\}
\label{th:Capacity of the ergodic DPC-PF}
The capacity of the channel in \eqref{eq:WritingOnFadingDirtChannelModelGeneral} is obtained as
\ea{
C=\max_{P_{U,X|S_R}} \ I(Y; U| \theta) - I(U;S_R),
\label{eq: capacityWritingOnFadingDirt QS}
}
\end{thm}
The result in Th. \ref{th:Capacity of the ergodic DPC-PF} holds for the general Gelf'and-Pinsker problem with partial channel state information at either the receiver or the transmitter but is stated in Th. \ref{th:Capacity of the ergodic DPC-PF} only for the model in \eqref{eq:WritingOnFadingDirtChannelModelGeneral}.
%

Equation \eqref{eq: capacityWritingOnFadingDirt QS} contains the auxiliary RV $U$ and is expressed as the maximization over $P_{U,X|S}$.
%
%
%
This expression is concave in $P_{U|S_R}$ for a fixed $P_{X|S_R, U}$ and convex in $P_{X|S_R,U}$ for a fixed $P_{U| S_R}$, which implies that $X$  can be chosen to be a deterministic function of $U$ and $S$.
Given the fact that \eqref{eq: capacityWritingOnFadingDirt QS} contains an auxiliary RV and given its convexity properties, it is not easy to obtain an expression of $C$ which depends solely on the channel parameters or to numerically approximate it.
For this reason, alternative inner and outer bounds have been derived in the literature.
%
%
In \cite{bennatan2008fading}, the RHS \eqref{eq: capacityWritingOnFadingDirt QS} is optimized for the case in which $U$ and $X$ are restricted to be Gaussian.
\begin{thm}{\bf Achievability with Gaussian signaling \cite[Sec. IV]{bennatan2008fading},\cite[Th. 1]{avner2010dirty}\\}
\label{th:linear assigment}
Let $\rho=(\rho_{xs},\rho_{us},\rho_{ux})$ and let $\Acal$ denote the region
\ea{
\Acal= \lcb \p{
|\rho_t|<1 \quad t \ \in\{xs,us,ux\} \\
1+2\rho_{xs} \rho_{us} - |\rho_{xs}|^2-|\rho_{us}|^2 - |\rho_{ux}|^2 =0
}\rcb
}
then, any distribution of $P_\theta$, the following rate is achievable
\ea{
R \leq \max_{\rho \in \Acal} \Ebb_{\theta} [ R_\Gamma(\rho, a) | \theta=a],
}
for
\ea{
& R_t(\rho,t) = \f 12 \log \lb (P+Q+ 2\Re\{\rho_{xs} t\} \sqrt{P Q}+1)(1-|\rho_{us}|^2) \rb  \nonumber \\
& -\f 12 \log \lb  P (1-|\rho_{ux}|^2)+Q (1-|\rho_{us}|^2) +  \rnone  \nonumber  \\
& \quad \quad  \quad \quad \lnone 2\Re\{t(\rho_{xs}- \rho_{ux} \rho_{us})\}\sqrt{P Q}+1\rb,
}

\end{thm}

\begin{IEEEproof}
The proof can be obtained from \cite[Th. 1]{avner2010dirty}  by noticing that the realization of $\phi$ which corresponds to the lowest achievable rate is
$\theta=\angle(X)$.
\end{IEEEproof}

An outer bound for the case where the fading is uniformly distributed among two values can be obtained from the ``carbon copying onto dirty paper'' \cite{khisti2007carbon} problem.
%

\begin{thm}{\bf Outer Bound for the Circularly Binomial Fading Dirt Channel \cite[Th. 5]{khisti2007carbon}\\}
\label{th:Outer Bound for the Quasi-Static Fading Dirt Channel}
The  capacity of the DPC-PF with the distribution of $\theta$ in \eqref{eq: distribution 2 values} is upper bounded as
\ea{
C  &  \leq  \f 12 \log(1+P)+\f 12 \log \lb 1+ (\sqrt{P}+\sqrt{Q})^2 \rb \nonumber \\
   & \quad \quad - \f 14 \log\lb 4 \sin(\Delta)^2 Q \rb.
\label{eq: capacityWritingOnFadingDirt QS}
}
\end{thm}
\begin{IEEEproof}
This result is a variation of the result in \cite[Th. 5]{khisti2007carbon} for $S_1 = e^{+\Delta i} S_R$ and $S_2=e^{-\Delta i} S_R$.
The full proof is provided in Appendix \ref{app:Outer Bound for the Quasi-Static Fading Dirt Channel}.
\end{IEEEproof}
The result \cite[Th. 5]{khisti2007carbon} was originally developed for the case in which the fading coefficient is fixed through successive channel uses.
%
%
The result in Th. \ref{th:Outer Bound for the Quasi-Static Fading Dirt Channel} is obtained by adapting the derivation in \cite[Th. 5]{khisti2007carbon} to the case of in which the fading changes at each channel use.

\section{Circular Binomial Phase Fading}
\label{sec:Two Phase Fading Values}

We begin by analyzing the scenario in which the phase fading takes only two values.
%
Since the uncertainty on the fading realization is limited, the encoder can efficiently cope with the interference through binning and Gaussian signaling  as in the channel without fading.
We begin by introducing the outer bound inspired by the ``carbon copying onto dirty paper'' of \cite{khisti2007carbon}.
The derivation is improved upon through a genie aided side information and by optimizing the outer bound over the power of the interference.
\begin{thm}{\bf Genie Aided Outer Bound\\}
\label{th: outer bound DPC-PF-U genie aided}
The  capacity of the DPC-PF with the distribution of $\theta$ in \eqref{eq: distribution 2 values} is upper bounded as
\ea{
& C \leq R^{\rm OUT-B} = \bigcup_{\gamma \in [0,1]}  \min_{[Q', \rho,c_Z,c_S] \in \Acal} \lb   \f 12 \log (T_1 T_2)+  \rnone \nonumber \\
& \quad  \quad \quad - \lnone \f 14 \log T_3 T_4 \rb +1,
\label{eq:outer bound QS DPC-PF}
}
with
$
\Acal = \lcb Q' \leq Q, \  \rho\in [-1,1],  c_+,c_-,c_S \in \Rbb  \rcb,
$
and for
\eas{
T_1 & = \lb 1+\gamma\sqrt{\f P {Q'}}\rb^2 Q'+1
\label{eq:terms H1 and H2 T1}  \\
& \quad \quad - \f{\lb (1+\gamma\sqrt{P/Q'})c_S Q' +c_{+}+c_{-} + 2 \rho c_{+} c_{-} \rb^2}{c_S^2 Q'+ c_{+}^2+c_{-}^2 + 2 \rho c_{+} c_{-} +1}
\nonumber \\
T_2 & =P(1-\gamma^2) + 1 - \f{(c_+ - c_-)^2(1-\rho)^2}{c_{+}^2+c_{-}^2 + 2 \rho c_{+} c_{-}+1}
\label{eq:terms H1 and H2 T2}\\
T_3 & = 4 \sin \lb \Delta \rb^2 Q' + 2(1-\rho)+
\label{eq:terms H1 and H2 T3}  \\
& \quad \quad - \f{ \lb 2 \sin \lb \Delta  \rb c_S Q'+(c_+ - c_-)(1-\rho) \rb^2} {c_S^2 Q'+ c_+^2 +c_-^2 +2 \rho c_+ c_- +1}
\nonumber \\
T_4 & = 2(1+ \rho) -   \f { ( c_+ + c_-)^2(1+ \rho)^2}{c_S^2 Q'+ c_+^2 +c_-^2 +2 \rho c_+ c_- +1},
\label{eq:terms H1 and H2 T4}
}{\label{eq:terms H1 and H2}}
\end{thm}

\begin{IEEEproof}
The proof follows the same line as \cite{khisti2007carbon} but with two further refinements.
The receiver is provided with a genie-aided side information which is obtained as a linear combination of the interference and the channel noise and an additional noise term, independent from all the other RV.
Moreover, the outer bound is optimized over the power of the interference in the range $[0,Q]$.
This is possible since the capacity of the channel increases as the power of the interference decreases.
%
%
%
The full proof is provided in Appendix \ref{app: outer bound QS DPC-PF}.
\end{IEEEproof}

The outer bound in Th. \ref{th: outer bound DPC-PF-U genie aided} is expressed as the optimization over multiple parameters and as the union over all $\gamma$.
%
%
We now derive a simpler outer bound which is expressed only as a function of the channel parameters.
%
%
\begin{lem}{\bf Simpler Outer Bound\\}
\label{lem:outer bound QS DPC-PF optimized 2 values}
If $\pi /4 \leq \Delta \leq \pi /2$,
the outer bound in Th. \ref{th: outer bound DPC-PF-U genie aided} can be further upper bounded as
\ea{
& C \leq R^{\rm OUT-APP-B}  =
\label{eq: outer bound QS DPC-PF optimized 2 values} \\
& \lcb \p{
 \log(P+1)+2
 &  \sin(\Delta)^2 Q \leq 1 \\
\f 34 \log(P+1)+2
& \sin(\Delta)^2 Q \geq P+1 \\
\f 12 \log(P+1)  & \\
\  + \f 12 \log\lb 1+(\sqrt{P}+\sin(\Delta)\sqrt{Q})^2  \rb & \\
\  - \f 14 \log(2 \sin(\Delta)^2 Q) +2
& 1 < \sin^2(\Delta) Q < P+1  \\
} \rnone. \nonumber
}
\end{lem}
\begin{IEEEproof}
The proof is similar to the proof in Th.  \ref{th: outer bound DPC-PF-U genie aided} but does not consider the genie aided side information.
The fundamental improvement from the proof in \cite{khisti2007carbon} is the optimization over the power of the interference.
While the capacity of the channel increases as $Q$ decreases, the outer bound is not monotonically decreasing in $Q$.
For this reason, optimizing the outer bound over $Q$ in the range $[0,Q]$ produces a tighter outer bound than \cite{khisti2007carbon}.
%
%
\end{IEEEproof}
We now derive an inner bound to the capacity region based on binning and Gaussian signaling as in the original DPC channel.
If the transmitter disregards the uncertainty over the phase fading and codes as in the DPC channel, it can pre-code successfully against the interference only half of the time on average.
Alternatively, the encoder can disregard the partial interference knowledge and transmit as if the interference were additional additive noise.
A scheme that combines the above two choices can be obtained by using two codewords to produce the channel input: one codeword is pre-coded against one realization of the interference while another codeword treats the interference as noise.
The performance of this scheme can then be optimized over the power allocated the two codewords.

\begin{thm}{\bf Interference as Noise and Binning Inner Bound\\}
\label{th:Interference as Noise + Binning Inner Bound}
The  capacity of the DPC-PF  with the distribution of $\theta$ in \eqref{eq: distribution 2 values} is lower bounded as
\ea{
& C \geq R^{\rm IN-B} =\f 12 \log \lb 1+\beb P  \rb + 
\label{eq:Interference as Noise + Binning Inner Bound} \\
& \ \  +  \f 12 \log \lb 1 + \f{\al \be P} {1 +\alb \be P+ \sin(\Delta)^2Q} \rb+ \nonumber \\
& \ \  +\f 14 \log \lb 1 + \alb \be P  \rb  + \nonumber \\
& \ \ +\f 14  \log \lb \max \lcb 1, \f {(\alb \be P+1)(\alb \be P+\sin(\Delta)^2 Q+1) }
{\alb \be P+2\sin(\Delta)^2 Q \alb \be P +\sin(\Delta)^2 Q+1}\rcb \rb,
\nonumber
}
for any $\al,\be \in [0,1]$ and $\alb=1-\al$, $\beb=1-\be$.
\end{thm}

\begin{IEEEproof}
%
On the real dimension, the interference sequence is always $\cos(\Delta) S_R$, so the encoder can pre-code  against this interference sequence as in the GP problem.
On the imaginary dimension, the interference is  $\sin(\Delta) S_R$ half of the time and $-\sin(\Delta) S_R$ the other half of the time.
On the imaginary dimension then, the encoder transmits two codewords, one which pre-codes against $\sin(\Delta) S_R$ while another codeword treats
$\sin(\theta)S_R$ as additional interference. Since $\sin(\theta)$ is uniformly distributed over $\{-\sin(\Delta), +\sin(\Delta)\}$, $\sin(\theta)S_R$ is Gaussian distributed.
This transmission scheme can be optimized over two parameters: $\be$, the ratio of the power used in the real versus imaginary dimension  and $\al$, the ratio of the power assigned to each codeword in the imaginary dimension.
The full proof is provided in Appendix \ref{app:Interference as Noise + Binning Inner Bound}.
\end{IEEEproof}
%
%
The inner bound in Th. \ref{th:Interference as Noise + Binning Inner Bound} is a function of two parameters, $\al$ and $\be$: a simpler inner bound expression can be obtained by carefully choosing the values of these two parameters.
%

\begin{lem}{\bf Simpler Inner Bound\\}
\label{lem:Approximate Inner Bound for Two Phase Fading Values}
The inner bound of Th. \ref{th:Interference as Noise + Binning Inner Bound} can be further lower bounded as
\ea{
& C \leq R^{\rm IN-APP-B} = 
\label{eq: optimize sectoring} \\
&\lcb \p{
\f 12 \log \lb 1+\f P 2\rb & \\
\quad  + \f 12 \log \lb 1  + \f P {2 + 2 \sin(\Delta)^2 Q^2} \rb                    & \sin(\Delta)^2 Q < 1  \\
\f 3 4 \log \lb 1+\f P 2\rb -1                                  & \sin(\Delta)^2 Q\geq  P+1   \\
\f 12 \log \lb 1+\f P 2\rb  & \\
\quad  + \f 12 \log \lb \f 12  + \f {P+2} {2\sin(\Delta)^2 Q} \rb & \\
\quad  + \f 14 \log (\sin(\Delta)^2 Q)-5/4                   & 1< \sin(\Delta)^2 Q < P+1 \\
}
\rnone
\nonumber
}
\end{lem}

\begin{IEEEproof}
The joint optimization of the two parameters $\al$ and $\be$ is quite hard, but fixing $\be$ makes it possible to optimize over $\al$ alone.
The expression in \eqref{eq: optimize sectoring} is obtained  by fixing $\be=1/2$ and optimizing the resulting expression over $\al$.
%
The full proof is provided  in Appendix \ref{app:Approximate Inner Bound for Two Phase Fading Values}.
\end{IEEEproof}

We now show that inner and outer bound in Lem. \ref{lem:Approximate Inner Bound for Two Phase Fading Values} and
Lem. \ref{lem:outer bound QS DPC-PF optimized 2 values} respectively are to within a finite gap for a subset of $\Delta$.
\begin{thm}{\bf Finite Gap between Inner an Outer Bounds \\}
\label{th:Constant Gap for Two Fading Values}
If $\pi / 4 \leq \Delta \leq \pi/2$, the inner bound in Th. \ref{th:Interference as Noise + Binning Inner Bound} and outer bound of Lem. \ref{lem:outer bound QS DPC-PF optimized 2 values} lie to within  constant gap of  3 bits/s/Hz.
\end{thm}

\begin{IEEEproof}
The inner and outer bound expressions are very similar and they can be compared for the case $\sin(\Delta)^2Q \geq P+1$ and the case $\sin(\Delta)^2Q < P+1$.
The full proof is provided in Appendix \ref{app:Constant Gap for Two Fading Values}.
\end{IEEEproof}

\section{Circular Uniform Phase Fading}
\label{sec:circular uniform Phase Fading}
We next focus on the case where the phase fading is uniformly distributed over the unitary circle.
For this scenario, both inner and outer bound in Sec. \ref{sec:Two Phase Fading Values} are no longer effective and new results are necessary to characterize capacity.
We begin by deriving an outer bound in which a genie provides the receiver with the phase of the channel input.

\begin{thm}{\bf Outer Bound\\}
\label{th:Outer Bound for Circularly Symmetric Phase Fading}
The  capacity of the DPC-PF with the distribution of $\theta$ in \eqref{eq: distirbution uniform} is upper bounded as
\ea{
C & \leq  R^{\rm OUT-C}  =\f 12 \log \lb 1+P \rb  + \nonumber \\
 & \f 12\log \lb 1+ P +Q +2 \sqrt{P Q}\rb - \f 12 \log\lb Q +1\rb  + 3/2 \label{eq:Outer Bound for Circularly Symmetric Phase Fading}.
 }
%
\end{thm}

\begin{IEEEproof}
The proof relies in adding a conditioning of a negative entropy term over the phase of the channel input.
This conditioning is used to divide the channel output in two components: one affected by the channel input, interference and noise, and another one only affected by the interference and noise.
%
This latter term can be easily evaluated since it is composed of a random mixture of iid Gaussian components.
The complete proof can be found in Appendix \ref{app:Outer Bound for Circularly Symmetric Phase Fading}.
\end{IEEEproof}

\begin{rem}
The outer bound in  Th. \ref{th:Outer Bound for Circularly Symmetric Phase Fading} can be tightened in the spirit of Th. \ref{th: outer bound DPC-PF-U genie aided} by providing a genie aided side information.
%
With this approach one obtains an outer bound expression in the spirit of \eqref{eq:outer bound QS DPC-PF} which can be optimized over the parameters in the side information.
%
\end{rem}

We next turn to the derivation of an inner bound for the circular uniform phase fading.
When the phase fading values are uniformly distritbuted over a large set, binning with Gaussian signaling provides only marginal advantages.
In particular, consider the Costa's dirty paper channel $Y=X+a S+Z$ with the assignment
$
U= X + \la S.
$
The achievable rate as a function of  $\la$ is
\ea{
R \leq \log\lb \f {P+a^2 Q+1} {1+\f Q P (\la^2+P(a-\la)^2)} \rb.
}
Let's assume that the transmitter has an incorrect estimate of $a$ and thus performs dirty paper coding for the gain $a+\ep$, instead of $a$.
In this case the attainable rate is
\ea{
R \leq \log\lb (1+P) \f {P+a^2 Q+1} { P+a^2 Q+1+Q P \ep^2} \rb,
}
so the achievable rate goes quickly to zero as the product $PQ \ep^2$ increases.
For this reason, binning with Gaussian signaling is not beneficial when the phase is circular uniform distributed, as even a small uncertainty over the exact channel realization drastically decreases the rate when $P$ and $Q$ are large.
%
Binning achieves capacity for this channel model, but only for Gaussian signaling it is possible to easily evaluate the achievable region.
Instead of focusing on determining a good assignment for $P_{U,X|S_R}$, we consider a different achievable scheme in which the imaginary dimension of the channel output is used to estimate the interference sequence and subtract it form the real dimension.
%
%
By combining the estimate of the interference dimension over the imaginary axes and the information transmitted over the real axe, the transmitter obtains an equivalent channel output which corresponds to a real fading channel.

\begin{thm}{\bf Real Transmission Inner Bound\\}
\label{th:Aligned Transmission in Uniform Noise}
The  capacity of the DPC-PF with the distribution of $\theta$ in \eqref{eq: distirbution uniform}  is lower bounded as
\ea{
C  \geq R^{\rm IN-U} & = \f 12 \log\lb 1  +Q+\al P\rb -\f 12 \log\lb 1+ Q\rb \nonumber \\
& \quad \quad +  \f 12 \log \lb  \alb P + 1 \rb -3,
\label{eq:Aligned Transmission in Uniform Noise}
}
for any $\al \in [0,1]$.
\end{thm}

\begin{IEEEproof}
The transmitter sends a codeword which threats the interference as noise on the imaginary dimension. After this codewords has been decoded, an estimate of the interference is produced.
More specifically, an estimate of $\sin(\theta^N) S_R^N$ is obtained from $\sin(\theta^N)  S_R^N+Z_I$.
Successively, the receiver estimates the real part of the channel input $X_R^N$ from
\eas{
\Yt^N & =\sin(\theta^N)Y_R^N - \cos(\theta^N)(\sin(\theta^N)  S_R^N+Z_I) \\
    & = \sin(\theta^N)X_R^N +  \sin(\theta^N) Z_R - \cos(\theta) Z_I \\
    & =\sin(\theta^N)X_R^N + \Zt
}
for $\Zt \sim \Ncal(0,1)$. The sequence $\Yt^N$ in therefore equivalent to the output of a real fading channel  with fading coefficient $\sin(\theta)$.
The complete proof is provided in Appendix \ref{app:Aligned Transmission in Uniform Noise}.
\end{IEEEproof}

We next show a gap between inner and outer bounds.

\begin{thm}{\bf Finite Gap between Inner and Outer Bounds \\}
\label{thm:Gap between Inner and Outer Bound for Uniform}
The gap between the inner bound in Th. \ref{th:Aligned Transmission in Uniform Noise} and in the outer bound in Th. \ref{th:Outer Bound for Circularly Symmetric Phase Fading} is at most $5.5$ bits/s/Hz.
\end{thm}
\begin{IEEEproof}
The difference between the expression in \eqref{eq:Outer Bound for Circularly Symmetric Phase Fading} and the
expression in \eqref{eq:Aligned Transmission in Uniform Noise} for $\al=1/2$ is $3+5/2= 5.5$ bits/s/Hz.
\end{IEEEproof}
The result of Th. \ref{thm:Gap between Inner and Outer Bound for Uniform} clearly implies that interference pre-cancellation is no longer useful when the transmitter has complete ignorance on the phase of the interference.
Instead, in the high interference regime, capacity is achieved by sacrificing half of the signal space to the estimation of the interference sequence.

\section{Numerical Simulations}
\label{sec:Numerical Simulations}

We now numerically simulate results of the previous sections to provide some further insight on the problem at hand.
\begin{figure}
\centering
\includegraphics[width=.55 \textwidth]{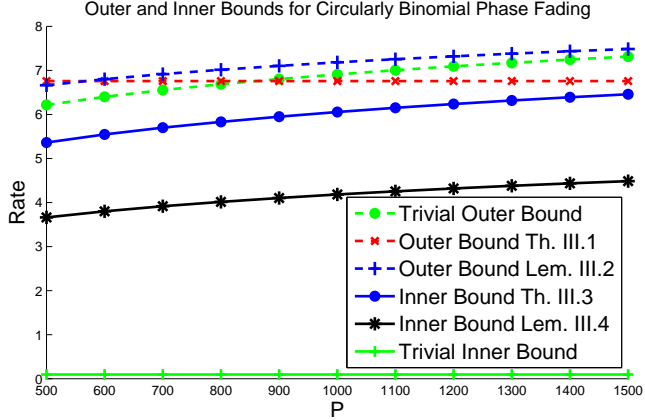}
\vspace{-.7 cm}
\caption{Inner and outer bound for the DPC-PF when the phase fading  is circular binomial for $P=[500 \ldots 1500]$ and $Q=10 P$.
Inner bounds are in solid lines while outer bounds are in dashed lines.}
\label{fig:binomial}
\vspace{-.5 cm}
\end{figure}
In Figure \ref{fig:binomial} we plot the inner bound in Th. \ref{th:Interference as Noise + Binning Inner Bound} and Lem. \ref{lem:Approximate Inner Bound for Two Phase Fading Values} and the outer bounds of Th. \ref{th: outer bound DPC-PF-U genie aided} in Lem. \ref{lem:outer bound QS DPC-PF optimized 2 values} for different values of $P$ and  $Q=10P$.
In the figure we also plot the trivial outer bound $R \leq \f12 \log(1+P)$, which is obtained by providing the interference sequence $S_R^N$ to the receiver, and the trivial inner bound $R \leq \f 12 \log(1+P+Q) + \f12 \log(1+Q)$, which is obtained by treating the interference as noise.
The constant gap result in Th. \ref{th:Constant Gap for Two Fading Values} is obtained comparing  Lem. \ref{lem:Approximate Inner Bound for Two Phase Fading Values} and Lem. \ref{lem:outer bound QS DPC-PF optimized 2 values}, but numerical simulations actually show a much smaller gap between more general inner and outer bounds.
\begin{figure}
\centering
\includegraphics[width=.5 \textwidth]{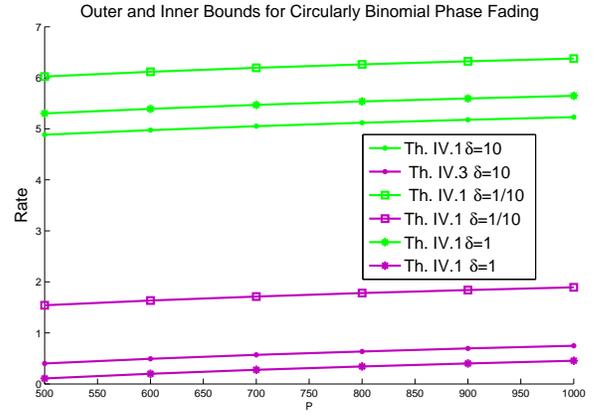}
\vspace{-.7 cm}
\caption{Inner and outer bound for the DPC-PF when the phase fading  is circular binomial for $P=[500 \ldots 1000]$ and $Q=10 P$, $Q=P$ and $Q=P/10$.}
\label{fig:uniform}
\vspace{-.5 cm}
\end{figure}
In Figure \ref{fig:uniform} we plot the inner and outer bound in Th. \ref{th:Aligned Transmission in Uniform Noise} and Th. \ref{th:Outer Bound for Circularly Symmetric Phase Fading} for increasing $P$ and different scaling $Q$: $Q=P/10$, $Q=P$ and $Q=10P$.
The distance between inner and outer bound is close to the gap result in Th. \ref{thm:Gap between Inner and Outer Bound for Uniform} and relatively insensitive to the ration between $P$ and $Q$.

\section{Conclusion}
\label{sec:Conclusion}

In this paper, we analyze the effect of phase fading in the classical Costa's dirty paper channel.
%
%
We consider a variation of the original setting in which the amplitude of the interfering sequence is known at the transmitter while its phase in known at the receiver.
Although derived in the literature, the capacity of this channel is hard to characterize in closed form or through numerical simulations.
We, therefore, derive the approximate characterization of capacity for the case in which the phase of the interference has a circular binomial and a circular uniform distribution.
%

\bibliographystyle{IEEEtran}
\bibliography{steBib1}

\newpage
\onecolumn
\appendix

\subsection{Proof of Th. \ref{th:Outer Bound for the Quasi-Static Fading Dirt Channel}}
\label{app:Outer Bound for the Quasi-Static Fading Dirt Channel}

From Fano's inequality we have
\eas{
R & \leq I(Y^N; W| \theta^N)  \\
  &  =   \sum_{j=1}^N  I(Y_j; W| \theta^N, Y^{j-1})  \\
  & \leq \sum_j \lb H(Y_j|\theta_j) - H(Y_j | W, \theta^N, Y^{j-1}) \rb.
}
For the  term $H(Y_j)$ we have
\eas{
H(Y_j) & = H(X_j+e^{i \theta_j} S_{Rj} + Z_j| \theta_j)   \\
&= \f 1 2 \lb H(X_j+e^{+i \Delta} S_{Rj} + Z_j) + H(X_j+e^{-i \Delta} S_{Rj} + Z_j)\rb \\ 
&= \f 1 2 \lb H(e^{+i \Delta} X_j +S_{Rj} + Z_j) + H(e^{-i \Delta}  X_j+ S_{Rj} + Z_j)\rb \\ 
&= \f 1 2 \lb H(\Re\{e^{+i \Delta} X_j\} +S_{Rj} + Z_{Rj}; \Im\{ e^{+i \Delta} X_j\} + Z_{Ij}) + H(\Re\{e^{-i \Delta} X_j\} +S_{Rj} + Z_{Rj}; \Im\{ e^{-i \Delta} X_j\} + Z_{Ij})\rb \\
& \leq \f 1 2 \lb H(\Re\{e^{+i \Delta} X_j\} +S_{Rj} + Z_{Rj}) + H (\Im\{ e^{+i \Delta} X_j\} + Z_{Ij}) + H(\Re\{e^{-i \Delta} X_j\} +S_{Rj} + Z_{Rj})+H( \Im\{ e^{-i \Delta} X_j\} + Z_{Ij})\rb \\
& \leq \f 12 \log \lb 1+P \rb  + \f 12 \log \lb 1+(\sqrt{P}+\sqrt{Q})^2  \rb.
\label{eq:last eq positive enropy}
}
Similarly, for the term $H(Y_j | W, \theta^N, Y^{j-1})$ we have 
\eas{
H(Y_j | W, \theta^N, Y^{j-1}) 
& = \f 1 2 \lb H(X_j+e^{+i \Delta} S_{Rj} + Z_j | W, \theta^{j-1},\theta_{j+1}^N, Y^{j-1}) + H(X_j+e^{-i \Delta} S_{Rj} + Z_j| W, \theta^{j-1},\theta_{j+1}^N, Y^{j-1})\rb \\
& \leq \f 12 H(X_j+e^{+i \Delta} S_{Rj} + Z_j, X_j+e^{-i \Delta} S_{Rj} + Z_j | W, \theta^{j-1},\theta_{j+1}^N, Y^{j-1}) \\
& = \f 12 H \lb \f 1 {\sqrt{2}}  2 \sin(\Delta) S_{Rj}  , \f 1 {\sqrt{2}} \lb 2 X_j+2 \cos(\Delta) S_{Rj} + 2 Z_j \rb | W, \theta^{j-1},\theta_{j+1}^N, Y^{j-1} \rb \\
& \leq \f 12 H \lb  \sqrt{2} \sin(\Delta) S_{Rj}   \rb +\f 12  H \lb  \f 1 {\sqrt{2}} \lb 2 X_j+2 \cos(\Delta) S_{Rj} + 2 Z_j \rb | W, \theta^{j-1},\theta_{j+1}^N, Y^{j-1}, S_{Rj} \rb \\
& \leq \f 12 H \lb  \sqrt{2} \sin(\Delta) S_{Rj}   \rb +\f 12  H \lb  \f 1 {\sqrt{2}} \lb 2 X_j+2 \cos(\Delta) S_{Rj} + 2 Z_j \rb | W, \theta^{j-1},\theta_{j+1}^N, Y^{j-1}, S_{R}^N,  \rb \\
& \leq \f 14 \log 2 \pi e Q+\f 12  H \lb  \f 1 {\sqrt{2}} \lb 2 X_j+2 \cos(\Delta) S_{Rj} + 2 Z_j \rb | W, \theta^{j-1},\theta_{j+1}^N, Y^{j-1}, S_{R}^N,  \rb \\
& \leq \f 14 \log 2 \pi e Q+\f 12  H \lb   \sqrt{2} Z_j \rb  \\
& \leq \f 14 \log 2 \pi e (4 Q) 
\label{eq:last eq negative enropy}\\
}{}

By combining the terms in  \eqref{eq:last eq positive enropy} and \eqref{eq:last eq negative enropy}, we obtain \eqref{eq: capacityWritingOnFadingDirt QS}

\subsection{Proof of Th. \ref{th: outer bound DPC-PF-U genie aided}}
\label{app: outer bound QS DPC-PF}

There are three components to the outer bound which we separately before the actual proof
\begin{itemize}
  \item the capacity of the channel is decreasing in $Q$, the power of the interference $S_R$,
  \item the correlation among the noise terms can be chosen as a function of $\theta_j$,
  \item the receiver is provided with a genie aided side information $U_j$.
\end{itemize}

\bigskip
\noindent
\emph{The capacity is decreasing in $Q$\\}
\noindent
Consider two sequences $S_{1R}^N$ and $S_{2R}^N$ for two independent $S_{mRj} \sim \iid \Ncal(0,Q_m), \ m \in \{1,2\}, \ j \in [1 \ldots N]$
with $Q=Q_1+Q_2$.
The interference sequence $S_{R}^N$ can be equivalently written as:
\ea{
S_{Rj}=S_{1Rj}+S_{2Rj}, \  j \in [1 \ldots N].
}
Providing $S_{2}^N$ to both the transmitter and receiver can only increase the capacity, since they can both disregard this extra information.
The capacity of the channel in which $S_{2}^N$ is provided to both encoder  and decoder follows in the class of channels studied in \cite[Th. 1]{cover2002duality}.
Capacity is thus obtained as
\eas{
C
&=  \max_{X,U| S_{2R},S_{R}} I(X+S_R e^{i \theta_j} + Z, \theta,  S_{2R} ; U) -I(U; S, S_{2R}) \\
&= \max_{X,U| S_{2R},S_{1R}} I(X+S_{1R} e^{i \theta_j} + Z, \theta, S_{2R} ; U) -I(U; S_{1R}, S_{2R}) \\
&=\max_{X,U| S_{2R},S_{1R}} I(X+S_{1R} e^{i \theta_j} + Z, \theta  ; U| S_{2R}) -I(U; S_{1R} | S_{2R}) \\
& \leq \max_{X,U| S_{2R},S_{1R}} I(X+S_{1R} e^{i \theta_j} + Z, \theta ; U,S_{2R}) -I(U,S_{2R}; S_{1R})
\label{eq: independent S1 S2}\\
& = \max_{X,\Ut| S_{2R},S_{1R}} I(X+S_{1R} e^{i \theta_j} + Z, \theta; \Ut) -I(\Ut; S_{1R}),
\label{eq: independent S1 S2 last}
}{\label{eq:genie}}
where, in \eqref{eq: independent S1 S2}, we have used the independence of $S_{1R}$ and $S_{2R}$ and for $\Ut=[U \ S_{2R}]$ in \eqref{eq: independent S1 S2 last}.
Since $S_{2R}$ does no longer appear in \eqref{eq: independent S1 S2 last}, we conclude that it can be dropped from the maximization.

The expression in \eqref{eq: independent S1 S2 last} corresponds to the capacity of the channel in \eqref{eq:WritingOnFadingDirtChannelModelGeneral} in which the interference has power $Q_1$ instead of $Q$.
This shows that the capacity of the channel in \eqref{eq:WritingOnFadingDirtChannelModelGeneral} is decreasing in  $Q$.

\bigskip
\noindent
\emph{Correlation among the noise terms\\}
\noindent
As in \cite{khisti2007carbon}, we notice that the joint distribution among the noise term $Z_j$ in \eqref{eq:WritingOnFadingDirtChannelModelGeneral}
can be chosen to depend on the realization of $\theta_j$, that is
\ea{
P_{Z_j,\theta_j} =P_{\theta_j} \lb  P_{Z_j | \theta_j=+\Delta} +   P_{Z_j | \theta_j=-\Delta} \rb
}
for two Gaussian RV with zero mean and unitary variance $Z_j|\theta_j=+\Delta$ and $Z_j|\theta_j=-\Delta$ that have any desired correlation.
This holds since the channel transition probability
\ea{
P_{\theta_j,Y_j}=P_{\theta_j}P_{Y_j|\theta_j}
}
is unaffected by the correlation between the RVs $Z_j|\theta_j=+\Delta$ and $   Z_j|\theta_j=-\Delta$.

For the sake of convenience we use the notation
\eas{
Z_{+j} & =Z_j|\theta_j=+\Delta \\
Z_{-j} & = Z_j|\theta_j=-\Delta.
}
and indicate with $\rho$ the correlation between the terms.
In general $\rho$  can be taken complex. In the following we focus on the case where $\rho$ is real, in which case imaginary parts are independent from the real parts.

\bigskip
\noindent
\emph{Genie aided side information\\}
\noindent
In the outer bound, the receiver is provided with a genie aided side information $U^N$ which is obtained as a linear combination of $S_{Rj}, Z_{+ j}$ and $Z_{-j}$, that is
\ea{
U_j= c_S S_{Rj} + c_{+Z} Z_{+ j} + c_{-Z} Z_{-j}  + \Zo_j,
\label{eq: def U}
}
for some iid $\Zo_i \sim \Ncal(0,1)$: $U^N$  and for some  $c_S, c_Z \ \in \Rbb$.

\bigskip

We now proceed with the derivation of the actual outer bound, starting from Fano's inequality:
\eas{
N(R - \ep_N)
& \leq I(Y^N, \theta^N ; W)  \\
& \leq H(Y^N, U^N; W  | \theta^N )\\
& =  H(Y^N ; W  | \theta^N, U^N)\\
& = \int \lb H(Y^N| U^N, \theta^N=\phi^N) - H(Y^N| W, U^N, \theta^N=\phi^N ) \rb \diff P_{\phi}^N,
\label{eq: integral}
}{}
where  $\phi^N \in  \{+\Delta, - \Delta \}^N$.

For the positive entropy term $H(Y^N| U^N, \theta^N=\phi^N)$ in \eqref{eq: integral} we have
\eas{
 H(Y^N| U^N, \theta^N=\phi^N)
&  = \sum_{j=1}^N H(Y_j| U^N, \theta^N=\phi^N, Y^{j-1}) \\
& \leq \sum_{j=1}^N H(Y_j| U_j, \theta_j=\phi_j)
\label{eq:CRE}\\
& \leq  N H(X_m+ e^{\phi_m} S_{Rm} + Z_m| U_m, \theta_m=\phi_m)
\label{eq:change to m}\\
& = \f N 2 \lb H(X_m+ e^{+\Delta} S_{Rm} + Z_{+m}|U_m) + H(X_m+ e^{-\Delta} S_{Rm} + Z_{-m}|U_m) \rb.
\label{eq:bound positive 2 values last}
}{\label{eq:bound positive 2 values}}
where \eqref{eq:CRE} follows from the conditioning reduces entropy property of the mutual information, \eqref{eq:change to m} is obtained by choosing the $m$ which maximizes the term $H(Y_j| U_j, \theta_j=\phi_j)$ over all $j=[1 \ldots N]$.
%
%

In the following we drop the subscript $m$ for ease of notation.
%
\eas{
H(X+ e^{+\Delta} S_{R} + Z_+| U)
& =H(e^{-\Delta}X+ S_R + e^{-\Delta} Z_+| U) \\
& =H(e^{-\Delta}X+ S_R + Z_+| U)
\label{eq: circular noise} \\
& =H(-\sin(\Delta) X+ S_R + Z_{R+}, \cos(\Delta) X+ Z_{I+}| U)
\label{eq: circular noise last}
}{}
where in \eqref{eq: circular noise} we have used the fact that the noise is circularly symmetric.
The choice of $X$  which maximizes  \eqref{eq: circular noise last} is of the form
\ea{
X_G = e^{\Delta} \lb \gamma \sqrt{\f P Q} S_R + \Xt_G \sqrt{P(1- \gamma^2)}  \rb
}
for some $\gamma \in [0,1]$ of some $\Xt_G \sim \Ccal(0,1)$.
With this choice we have
\eas{
H(-\sin(\Delta) X+ S_R + Z_{R+}, \cos(\Delta) X+ Z_{I+}| U)
& = H \lb \lb 1+\gamma \sqrt{\f P Q} \rb  S_R + Z_{R+}| U\rb + H\lb  \sqrt{P(1-\gamma^2)} \Xt + Z_{I+} | U \rb \\
& = \f 12 \log(2 \pi e T_1 ) + \f 12 \log(2 \pi e T_2)
}
where $T_1$ is obtained as
\eas{
T_1 & = \var[ (1+\gamma \sqrt{P/Q}) S_R + Z_{R+} | U] \\
& = \var \lsb (1+\gamma \sqrt{P/Q}) S_R + Z_{R+} | c_S S_R+ c_+ Z_{R+} + c_- Z_{R-}) + \Zt_R \rsb
\label{eq:imm perp real} \\
&=\eqref{eq:terms H1 and H2 T1}
}
where \ref{eq:imm perp real} follows from the fact that $\rho$ is positive and real and imaginary parts are independent.
The term $T_2$ is obtained as
\eas{
T_2 & = \var\lsb \sqrt{P(1-\gamma^2)} \Xt_G  + Z_{I+} | U \rsb \\
& = \var\lsb \sqrt{P(1-\gamma^2)} \Xt_G + Z_{I+} | c_Z (Z_{I+} + Z_{I-}) + \Zt_{I}  \rsb \\
& = \eqref{eq:terms H1 and H2 T2}
}
\medskip

The term $H(X_m+ e^{-\Delta} S_{Rm} + Z_{-m}|U_m)$ is bounded in an analogous manner to yield the same expression.

\bigskip

Let's now focus on the negative entropy term:
\eas{
& - \int H(Y^N| W, U^N, \theta^N=\phi^N) \diff P_{\phi}^N\\
& =    - \f 12 \int \lb H(Y^N| W, U^N, \theta^N=\phi^N) + H(Y^N| W, U^N, \theta^N=-\phi^N)  \rb \diff P_{\phi}^N
\label{eq: double entropy 2 values}\\
& \leq - \f 12 \int H(X^N+ e^{i \phi^N}S_R^N + Z_{+\phi}^N; X^N +e^{-i \phi^N} S_R^N + Z_{-\phi}^N|W, U^N ) \diff {P_\phi^N}
\label{eq:two entropy terms 1} \\
&= - \f 12 \int \lb H(2 i \sin(\phi^N)S_R^N+ Z_{+\phi}^N-Z_{-\phi}^N, 2 X^N + 2 \cos(\phi^N) S_R + Z_{+\phi}^N-Z_{-\phi}^N | U^N, W)   \rb \diff P_{\phi}^N  - N \log 2
\label{eq:two entropy terms 2} \\
& \leq - \f 12 \int \lb H(2 \sin(\phi^N)S_R^N+ Z_{+\phi}^N-Z_{-\phi}^N| U^N)  \rnone \\
& \quad \quad \quad \quad  \lnone -H( 2 X^N + 2 \cos(\phi^N) S_R + Z_{+\phi}^N-Z_{-\phi}^N | U^N, W,2 \sin(\phi^N)S_R^N+ Z_{+\phi}^N+Z_{-\phi}^N)   \rb \diff P_{\phi}^N  - N  \log 2 \\
& \leq - \f 12 \int \lb N H(2 i \sin(\phi_j)S_{Rj}+ Z_{+\phi j}-Z_{-\phi j}| U_j)-H( 2 X_j + \cos(\phi_j) S_{Rj} + Z_{+\phi_j}-Z_{-\phi_j} | U^N, W, S_R^N)   \rb  \diff P_{\phi}^N -  N  \log 2
\label{eq:two entropy terms 3}\\
&= - \f N 2 \int \lb H(2 \sin(\phi_j)S_{Rj}+ Z_{+\phi j}-Z_{-\phi j}| U_j)-H( Z_{+\phi j}+Z_{-\phi j} | U_j)   \rb  \diff P_{\phi_j} - N  \log 2\\
&= - \f N 2 \int \lb H(2 \sin(\Delta)S_{R}+ Z_{+}-Z_{-\phi}| U)-H( Z_{+\phi}+Z_{-\phi} | U)   \rb  \diff P_{\phi} - N \log 2\\
\label{eq:two entropy terms 4}\\
& = -\f N 2 \log(2 \pi e T_3) - \f N 2 \log (2 \pi e T_4) - N \log 2,
}{}
where \eqref{eq: double entropy 2 values} is obtained by paring each sequence $\phi^N$ with the complement sequence $-\phi^N$, in which each $+\Delta$ is replaced by a $-\Delta$ and each $-\Delta$ by a $+\Delta$.
In \eqref{eq:two entropy terms 1} we define $Z_{\phi}^N$ as the sequence of noise terms associated with the sequence of phase fading values $\phi^N$.
The passage in \eqref{eq:two entropy terms 2} is obtained with the transformation
\ea{
H(U_1,U_2)= H(U_1+U_2,U_1-U_2) - \f 12.
}
In \eqref{eq:two entropy terms 3} we have used the fact that the state and the noise are iid for the first term and the ``conditioning reduces entropy''
property of the entropy for the second term.
In \eqref{eq:two entropy terms 4} we have used the fact that the expression on longer depends on $j$ or $\theta_j$.

The terms $T_3$ and $T_4$ are obtained as
\eas{
T_3 & = \var[2 i \sin(\Delta)S_{R}+ Z_{+}-Z_{-}| U]= \eqref{eq:terms H1 and H2 T3}\\
T_4 & = \var [Z_{+}-Z_{-}| U] = \eqref{eq:terms H1 and H2 T4}
}{}

%
%
%
\subsection{Proof of Lem. \ref{lem:outer bound QS DPC-PF optimized 2 values}}
\label{app:outer bound QS DPC-PF optimized 2 values}

The proof we consider is a variation of the proof of Th. \ref{th: outer bound DPC-PF-U genie aided} in App. \ref{app: outer bound QS DPC-PF}.

\medskip
We consider the case where no side information is provided at the receiver but still optimize over the correlation between the noise terms and over the power of the interference.

\bigskip
For the positive entropy term in $H(Y^N| U^N, \theta^N=\phi^N)=H(Y^N| \theta^N=\phi^N)$ in \eqref{eq: integral} we have
\eas{
 H(Y^N| \theta^N=\phi^N)
& \leq  N H(X_j+ e^{\phi_j} S_{Rj} + Z_j|\theta_j=\phi_j) \\
& \leq \f N 2 \lb H(X_j+ e^{\theta_S} S_{Rj} + Z_j) + H(X_j+ e^{-\theta_S} S_{Rj} + Z_j) \rb.
\label{eq:bound positive 2 values last}
}{\label{eq:bound positive 2 values}}

When $\pi/4 \leq \Delta \leq \pi/2 $, we have that  $\cos(\Delta) \leq \sin(\Delta) $, we can write
\eas{
&H(X+ e^{\theta} S_{R} + Z) \\
& = H(X_I + \sin(\theta) S_{R} + Z_I; X_R + \cos(\theta) S_{R} + Z_R)  \\
& \leq \f 12 \log \lb 1 + P + \sin(\Delta)^2Q + 2 \sqrt{PQ}\rb + H(X_R + \cos(\theta) S_{R} + Z_R | X_I + \sin(\theta) S_{R} + Z_I) \\
& = \f 12 \log \lb 1 + P + \sin(\Delta)^2 Q + 2 \sin(\Delta)\sqrt{PQ}\rb + H(X_R + \cos(\theta) S_{R} + Z_R | X_I + \sin(\theta) S_{R} + Z_I) \\
& = \f 12 \log \lb 1 + P + \sin(\Delta)^2 Q + 2 \sin(\Delta)\sqrt{PQ}\rb + H\lb X_R+\f {\cos(\theta)}{\sin(\theta)}X_I  + Z_R + \f {\cos(\theta)}{\sin(\theta)} Z_I | X_I + \sin(\theta) S_{R} + Z_I\rb \\
& \leq \f 12 \log \lb 1 + P + \sin(\Delta)^2 Q + 2 \sin(\Delta)\sqrt{PQ}\rb + H\lb X_R+\f {\cos(\theta)}{\sin(\theta)}X_I  + Z_R + \f {\cos(\theta)}{\sin(\theta)} Z_I \rb \\
&  \leq \f 12 \log \lb 1 + P + \sin(\Delta)^2 Q + 2 \sin(\Delta)\sqrt{PQ}\rb + \f 12 \log \lb 4 P + 2 \rb, \\
&  \leq \f 12 \log \lb 1 + P + \sin(\Delta)^2 Q + 2 \sin(\Delta)\sqrt{PQ}\rb + \f 12 \lb P + 1 \rb +1,
}

For the negative entropy term $- H(Y^N| W, U^N, \theta^N=\phi^N )$ in \eqref{eq: integral} we set $U^N=\Zo^N$.
The overall outer bound is now
\ea{
R^{\rm OUT-B}=\f 12 \log (1+P) +\min_{\rho,Q'} \lb \f12 \log(1+P+ \sin(\Delta)^2 Q'+2\sqrt{PQ'})-\f 14 \log((2\sin(\Delta)^2 Q'+1-\rho)(1+\rho)) \rb+2.
\label{eq:entropy minus 2}
}
We can now optimize this expression in \eqref{eq:entropy minus 2} over $\rho$ and $Q'$.
The optimal assignment for $\rho$ is
\ea{
\rho^*=\min\{1,\sin(\Delta)^2Q\},
}
and, for $\rho=\rho^*$,  the optimal value of $Q'$ is
\ea{
\sin(\Delta)^2 Q^*=\min\{P+1,\sin(\Delta)^2 Q\}.
}
With this assignment, we obtain the bound in \eqref{eq: outer bound QS DPC-PF optimized 2 values}.

\subsection{Proof of Th. \ref{th:Interference as Noise + Binning Inner Bound}}
\label{app:Interference as Noise + Binning Inner Bound}

We analyze the performance on the real and the imaginary dimension separately.
The transmitter can decide how to assign power in the two dimensions:  in the following we assume that
\ea{
E[X_I^2]=\be P  \\
E[X_R^2]=\beb P,
}
for $\beb=1-\be$.

\bigskip
\noindent
\emph{Real Dimension\\}
\noindent
On the real dimension, the interference sequence is always $\cos(\Delta)S_R^N$ and therefore in is possible to code as in the classical GP problem \cite{GelFandPinskerClassic}.
In particular,  from the classical ``writing on dirty paper'' result \cite{costa1983writing} we have that  the assignment
\ea{
X_R, & \sim \Ncal(0, \beb P) \\
U_R & = X_R + \f {\beb P}{\beb P+1} \cos(\Delta) S_R,
}
attains the rate achievable
\ea{
R_R=\f 12 \log(1+ \beb P),
}
on the real dimension.

\bigskip
\noindent
\emph{Imaginary Dimension\\}
\noindent
The imaginary channel input, $X_I$, is composed of two codewords:
\begin{itemize}
  \item a first codeword e $X_{IN}^N$ ($I$ for ``\emph{Imaginary}'', $N$ as in ``\emph{interference as Noise}'') which treats the interference as noise while
  \item a second codeword $X_{IP}$ ($I$ for ``\emph{Imaginary}'', $P$ as in ``\emph{Pre-coded against the interference}'')  is pre-coded against the sequence $+\sin(\Delta)S_R^N$.   This pre-coding offers full interference pre-cancellation half of the time while only partial interference pre-coding the rest of the time.
\end{itemize}

%
%
%
%
The codeword $X_{IN}$ is decoded first  and removed from the channel output and, successively, the codeword $X_{IP}^N$:
this strategy attains the rate
\eas{
R_{IN} & \leq  I(Y_I;X_{IN}|\theta) \\
R_{IP} & \leq  I(Y_I, \theta; U_{IP}|X_{IN}) - I(U_{IP}; \sin(\theta)S_R),
\label{eq:achievable rate noise and pre-coding}
}
with  $R_I=R_{IN}+R_{IP}$.
We consider, in particular, the assignment
\eas{
X_{IN} & \sim \Ncal(0, \al P/2) \\
X_{IP} & \sim \Ncal(0, \alb P/2) \\
&\al \in [0,1], \ \alb=1-\al\\
X_I    &=X_{IN}+X_{IP} \\
U_{IP} &= X_{IP}+ \f{\alb P}{\alb P+1} S_R.
}
This assignment attains
\ea{
R_{IN} = \f12 \log \lb 1+ \f {\al \be P } {1 + \alb \be P + \sin(\Delta)^2 Q}\rb,
\label{eq:immaginary startegy 2 values noise}
}
and
\eas{
R_{IP}
& =  \f 12 H(U| Y, \theta=\Delta)  + \f 12 H(U| Y, \theta=-\Delta) + H(X_{IP}) \\
&=\f 14 \log \lb 1 + \alb \be P  \rb +\f 14  \log \lb \min \lcb 1, \f {(\alb \be P+1)(\alb \be P+\sin(\Delta)^2 Q+1) }
{\alb \be P+2\sin(\Delta)^2 Q \alb \be P +\sin{\Delta}^2 Q+1}\rcb \rb.
}{\label{eq:immaginary startegy 2 values precoded}}

\subsection{Proof of Lem. \ref{lem:Approximate Inner Bound for Two Phase Fading Values}}
\label{app:Approximate Inner Bound for Two Phase Fading Values}

Consider the inner bound of Th. \ref{th:Interference as Noise + Binning Inner Bound} for $\be=1/2$  and disregard the last term in \eqref{eq:Interference as Noise + Binning Inner Bound}.
The inner bound is then further lower bounded by
\ea{
R^{\rm IN-B} \geq \f 12 \log \lb 1+\f P 2 \rb  + \f 12 \log \lb 1 + \f{\al P} {2 +\alb P+ 2 \sin^2(\Delta)^2Q} \rb+\f 14 \log \lb 1 + \f {\alb P } 2 \rb.
\label{eq:Time Sharing + Binning Inner Bound app}
}
The derivative of RHS of \eqref{eq:Time Sharing + Binning Inner Bound app} in $\al$ is
\ea{
D= - \f {P}2 \f{1+\al P-\sin(\Delta)^2 Q} {(1+\al P)(1+\al P+\sin(\Delta)^2 Q)},
}
therefore, if $\sin(\Delta)^2 Q>P+1$, then $\al=1$ is optimal.
If $0 \leq \sin(\Delta)^2 Q-1 < P$, the optimal $\al$ is $\f{\sin(\Delta)^2 Q-1} P$ while, if $\sin(\Delta)^2 Q < 1$, the optimal value is $\al=0$.

\subsection{Proof of Th. \ref{th:Constant Gap for Two Fading Values}}
\label{app:Constant Gap for Two Fading Values}

Consider first the case  $\sin(\Delta)^2 Q<1$: in this case by treating the interference as noise in the imaginary dimension we attain
\ea{
R_I & = \f 12 \log \lb 1 + \f  {P}{2+ 2 \sin(\Delta)^2 Q}\rb \\
    & \geq  \f 14 \log \lb \f 1 4 + \f  P 4 \rb  \\
    & \geq \f 12 \log(1+P) -1,
}
while, using Costa pre-coding on the real axe, we attain
\ea{
R_R = \f 12 \log(1+P/2) \geq \f 12 \log(1+P)-\f 12.
}
The gap between inner and outer bound when $\sin(\Delta)^2 Q \leq 1$ is therefore $1.5$ bits/s/Hz
Let's now compare inner and outer bound expression for $\sin(\Delta)^2 Q > 1$ by considering the case $\sin(\Delta)^2 Q \geq P+1$ and
$\sin(\Delta)^2 Q < P+1$.

By comparing the outer bound in \eqref{eq: outer bound QS DPC-PF optimized 2 values} and the inner bound in \eqref{eq: optimize sectoring}
for the case $\sin(\Delta)^2 Q>1$ and $\sin(\Delta)^2 Q \geq P+1$ we see that the two bounds differ by 3 bits/s/Hz.

For the case $\sin(\Delta)^2 Q < P+1$ we have
\eas{
R^{\rm OUT-APP-B} - R^{\rm IN-APP-B} & = \f 12 \log \lb \f{2 P + 2}{P+1}\rb+  \f 12 \log \lb \f{3+4 P}{\f 12 + \f{P+1} {\sin(\Delta)^2 Q}} \rb   - \f 12 \log (\sin(\Delta)^2 Q)    \\
& = \f 12 + \f 12 \log\lb \f{3+4 P}{ \f {\sin(\Delta)^2 Q} 2 +P+1}\rb \leq \f 3 2,
}and we see that the distance between inner and outer bound is at most 3 bits/s/Hz.

\subsection{Proof of Th. \ref{th:Outer Bound for Circularly Symmetric Phase Fading}}
\label{app:Outer Bound for Circularly Symmetric Phase Fading}

By applying Fano's inequality we obtain
\eas{
N (R -\ep_N) &  \leq I(Y^N; W | \theta^N) \\
& \leq H(Y^N| \theta^N) - H(Y^N| W, \theta^N).
\label{eq:fano circular uniform last}
}{\label{eq:fano circular uniform}}
For the positive entropy term in \eqref{eq:fano circular uniform last} we have
\eas{
H(Y^N|\theta^N)
& = \sum_j H(Y_j|\theta^N, Y^{j-1}) \\
& \leq \sum_j H(Y_j|\theta_j) \\
& \leq N H(Y_m|\theta_m)
\label{eq:HY passage 1} \\
& = N H(X  + S_R e^{i \theta}+ Z| \theta) \\
& = N H(X e^{-i \theta} + S_R + Z| \theta)
\label{eq:HY passage 2} \\
&  = N H( \Im \{ e^{i \theta} X \}+Z_I ; \Re \{ e^{i \theta} X \} + S_R + Z_R)  \\
& \leq \f N 2 \log \lb P  +1 \rb +\f N 2 \log \lb 1+ P +Q + 2\sqrt{PQ} \rb,
& \leq \f N 2 \log 2 \pi e \lb P  +1 \rb +\f N 2 \log  2 \pi e\lb 1+ P +Q \rb+\f 12,
\label{eq:HY last passage}
}

where \eqref{eq:HY passage 1} follows from choosing $m$ that maximizes $H(Y_m|\theta_m)$ .
In the following passages the index $m$ is dropped for convenience.
In \eqref{eq:HY passage 2} we have used the fact that the noise is circularly symmetric and thus rotations do not affect its distribution.
In \eqref{eq:HY last passage} we have used the fact that
$$
P + Q + 2 \sqrt{2 PQ} +1 \leq 2 P + 2 Q + 2
$$

%

For the term $H(Y^N| W, \theta^N)$  we provide the phase of the channel input $X^N$ as a genie aided side information to the receiver,
Let
\ea{
\angle X^N=\psi^N
}
so that we can write
\eas{
- H(Y^N| W, \theta^N) & \leq - H(Y^N| W, \phi^N,\psi^N)  \\
 & \leq - H( |X|^N e^{i \psi^N} + S_R e^{i \theta^N} + Z^N| W, \phi^N,\psi^N ) \\
 & = - H( |X|^N  + e^{i (\theta-\phi)^N} S_R^N + Z^N| W, (\phi-\psi)^N) \\
 & = - H( \sin((\phi-\psi)^N) S_R^N  + Z_I^N, |X|^N  + \cos((\phi-\psi)^N) S_R^N  + Z_R^N,  | W, (\phi-\psi)^N)  \\
 & \leq - H(\sin((\phi-\psi)^N) S_R^N  + Z_I^N |  (\phi-\psi)^N)  - H(|X|^N  + \cos((\phi-\psi)^N) S_R^N  + Z_R^N,  | W, (\phi-\psi)^N, S_R^N)   \\
}
Note now, that regardless of the distribution of $\psi^N$,  $(\phi-\psi)^N$ is iid and uniformly distributed over $[0,\pi)^N$ since $\phi^N$ is iid and uniformly distributed over $[0,\pi)^N$.
For this reason we can write
\ea{
- H(Y^N| W, \theta^N)  & \leq - N H( \sin(\phi_j-\psi_j) S_R +Z_I |  \theta_j-\psi_j) - H(|X|^N  + \cos((\phi-\psi)^N) S_R^N  + Z_R^N,  | W, (\phi-\psi)^N, S_R^N)   \\
& \leq  - N H( \sin(\phi_j-\psi_j) S_R +Z_I |  \phi_j-\psi_j) - \f N 2   \log \lb 2 \pi e \f 1 2 \rb \\
&= - \f 1 {2 \pi } \int_0^{2 \pi}  \f 12 \log 2 \pi e \lb \sin(t)^2 Q + 1 \rb \diff t -  \f N 2   \log(2 \pi e)\\
&= - \f 1 {2 \pi } \int_0^{2 \pi}  \f 12 \log \lb \sin(t)^2 Q + 1 \rb \diff t -  N  \log(2 \pi e) .
}
We now use the fact that
\eas{
& \sin(t) \geq \f 2 \pi  t, \quad t \in [0, \pi/2] \\
& \log (\sin(t)^2 Q+1 ) \geq \log \lb  \f 4 {\pi^2}t^2 Q+1  \rb, \quad t \in [0, \pi/2]
}{\label{eq:inequality sin}}
so to obtain
\eas{
- \f 1 {2 \pi } \int_0^{2 \pi}  \f 12 \log \lb \sin(\phi)^2 Q + 1 \rb
& \leq  -\f 12 \log(Q+1) +1 -\f{\pi \arctan(\sqrt{Q})}{\sqrt{Q}} \\
& \leq -\f 12 \log(Q+1) +1.
}{\label{eq:inequality sin capacity expression}}
This concludes the proof.

\subsection{Proof of Th. \ref{th:Outer Bound for Circularly Symmetric Phase Fading}}
\label{app:Outer Bound for Circularly Symmetric Phase Fading}
By applying Fano's inequality we obtain
\eas{
N (R -\ep_N) &  \leq I(Y^N; W | \theta^N) \\
& \leq H(Y^N| \theta^N) - H(Y^N| W, \theta^N).
\label{eq:fano circular uniform last}
}{\label{eq:fano circular uniform}}
For the positive entropy term in \eqref{eq:fano circular uniform last} we have
\eas{
H(Y^N|\theta^N)
& \leq N H(Y_j|\theta) \\
& \leq N H(X e^{i \theta} + S_R+ Z) \\
&  = N H( \Im \{ e^{i \theta} X \}+Z_I ; \Re \{ e^{i \theta} X \} + S_R + Z_R)  \\
& = \f N 2 \log \lb P  +1 \rb +\f N 2 \log \lb 1+ P +Q + 2\sqrt{PQ} \rb
\label{eq:HY last passage}
}
%

To bound the negative entropy term $-H(Y^N| W, \theta^N)$  in \eqref{eq:fano circular uniform last} we introduce the a conditioning over the phase of the channel input $X^N$.
For ease of notation let
\ea{
\angle X^N=\psi^N
}
so that we can write
\eas{
- H(Y^N| W, \theta^N) & \leq - H(Y^N| W, \phi^N,\psi^N)  \\
 & \leq - H( |X|^N e^{i \psi^N} + S_R e^{i \theta^N} + Z^N| W, \phi^N,\psi^N ) \\
 & = - H( |X|^N  + e^{i (\theta-\phi)^N} S_R^N + Z^N| W, (\phi-\psi)^N) \\
 & = - H( \sin((\phi-\psi)^N) S_R^N  + Z_I^N, |X|^N  + \cos((\phi-\psi)^N) S_R^N  + Z_R^N,  | W, (\phi-\psi)^N)  \\
}
Note now, that regardless of the distribution of $\psi^N$,  $ \vartheta^N=(\phi-\psi)^N$ is iid and uniformly distributed over $[0,\pi)^N$ since $\phi^N$ is iid and uniformly distributed over $[0,\pi)^N$.
For this reason we can write
\eas{
- H(Y^N| W, \theta^N)  & \leq - H( \sin(\vartheta^N) S_R^N  + Z_I^N, |X|^N  + \cos(\vartheta^N) S_R^N  + Z_R^N,  | W, \vartheta^N)  \\
&  = - H( \sin(\vartheta^N) S_R^N +Z_I^N |  \vartheta^N) - H(|X|^N  + \cos(\vartheta^N) S_R^N  + Z_R^N| W, \vartheta^N, \sin(\vartheta^N) S_R^N +Z_I^N)   \\
& \leq - H( \sin(\vartheta^N) S_R^N +Z_I^N |  \vartheta^N) - H(|X|^N  + \cos(\vartheta^N) S_R^N  + Z_R^N| W, \vartheta^N, S_R^N)   \\
& = - N H( \sin(\vartheta) S_R +Z_I |  \vartheta) - H(Z_R^N| W, \vartheta^N, S_R^N)   \\
& \leq  - N H( \sin(\vartheta) S_R +Z_I | \vartheta) - \f N 2  \log \lb 2 \pi e \rb \\
&= - \f 1 {2 \pi } \int_0^{2 \pi}  \f 12 \log \lb \sin(t)^2 Q + 1 \rb \diff t -  \f N 2   \log(2 \pi e)\\
&= - \f 2 { \pi } \int_0^{\pi/2}  \f 12 \log \lb \sin(t)^2 Q + 1 \rb \diff t -  \f N 2   \log(2 \pi e).
}
We now use the fact that
\eas{
& \sin(t) \geq \f 2 \pi  t, \quad t \in [0, \pi/2] \\
& \log (\sin(t)^2 Q+1 ) \geq \log \lb  \f 4 {\pi^2}t^2 Q+1  \rb, \quad t \in [-\pi/2, \pi/2]
}{\label{eq:inequality sin}}
so to obtain
\eas{
- \f 1 {2 \pi } \int_0^{2 \pi}  \f 12 \log \lb \sin(\phi)^2 Q + 1 \rb
& \leq  -\f 12 \log(Q+1) +1 -\f{\pi \arctan(\sqrt{Q})}{\sqrt{Q}} \\
& \leq -\f 12 \log(Q+1) +1.
}{\label{eq:inequality sin capacity expression}}
Combining  \eqref{eq:HY last passage} and \eqref{eq:inequality sin capacity expression} we obtain the expression in \eqref{eq:Outer Bound for Circularly Symmetric Phase Fading}.
%
%
%
%

\subsection{Proof of Th. \ref{th:Aligned Transmission in Uniform Noise}}
\label{app:Aligned Transmission in Uniform Noise}

The transmission scheme can be described as:
\begin{itemize}
  \item a first codeword e $X_{IN}^N$ ($I$ for ``\emph{Imaginary}'', $N$ as in ``\emph{interference as Noise}'') which treats the interference $e^{i \theta}S_R$ as noise while
  \item a second codeword, $X_{RC}^N$, ( $R$ for ``\emph{Real}'' and   for ``interference \emph{Cancellation}'') is transmitted only on the real axe and is decoded after $X_{IN}^N$ and  after having subtracted from the real channel output the estimate of the interfere $S_R$ obtained from the imaginary channel output.
\end{itemize}
%
%
%
%
%
More specifically, consider the following transmission scheme:

\textbf{Codebook Generation:}
The message $W$ is split into two sub-messages $W_N$ and $W_C$.
The sub-message $W_N$ is encoded in $X_{IN}^N$, $I$ for \emph{imaginary} and $N$ for ``treating the interference as \emph{Noise}'' and  the codebook for $X_{IN}^N$ is generated by drawing $2^{N R_IN}$ sequences of length $N$ with iid draws from the distribution $\Ncal(0,\al P)$.
The sub-message $W_C$ is encoded in $X_{RC}^{N}$  which is generated by iid drawings $2^{N R_{RC}}$ sequences of length $N$ with iid draws from the distribution $\Ncal(0,\alb P)$.
Each codeword is indexed as $X_{RC}^N(j),  \ j \in [1 \ldots 2^{N R_{RC}}-1]$.

\textbf{Encoding:}
Each channel input $X_j$ is obtained as
\ea{
X^N=X_{RC}^N+i X_{IN}^N,
}
which is a complex Gaussian (although not circularly symmetric) with covariance $P$.

\textbf{Decoding:}
The decoder first decodes $X_N^N$: this can be done as long as
\eas{
N R_{IN} & \leq I(Y_I^N; X_{IN}^N| \theta)\\
    & \leq N I(X_I+ \sin(\theta_j) S_R + Z_I ;X_{IN}|\theta_j),
}
which can yields
\ea{
R_{IN} & = \int \f 12 \log\lb 1 + \f{\al P}{1+\sin(t) Q } \rb  \diff P_{\theta}(t).
}
A close form evaluation of $R_{IN}$ is possible only through inequality similar to \eqref{eq:inequality sin} since
\eas{
\sin(t) & \geq  t, \ t \in [0 \ldots \pi/2]\\
\log\lb 1 + \f{\al P}{1+\sin(t)^2 Q } \rb   & \geq  \log\lb 1 + \f{\al P}{1+t^2 Q } \rb,
}
so that we obtain
\eas{
R_{IN} & \geq 4  \f  1 {2 \pi} \int_0^{\f \pi 2} \f 12 \log\lb 1 + \f{\al P}{1+t^2 Q } \rb \diff t\\
&=\f 12 \log\lb 1+\f  {\al P} {\f {\pi^2} 4 Q  +1} \rb +  \f {\arctan \lb  \sqrt{ \f{ \pi^2/4 Q }{P+1}} \rb  } {\f{ \pi^2/4 Q }{P+1} } -  \\
& \quad \quad - \f {\arctan \lb \sqrt{\pi^2/4 Q} \rb} {\sqrt{\pi^2/4 Q}}
\label{eq:arctan term}\\
}
since \eqref{eq:arctan term} is monotonically decreasing in $Q$ and
\ea{
\lim_{Q \goes 0} \f {\arctan \lb \sqrt{\pi^2/4 Q} \rb} {\sqrt{\pi^2/4 Q}} =2
}
we conclude that
\ea{
R_{IN}
& \geq \f 12 \log\lb 1+\f  {\al P} {4Q  +1}\rb -2 \\
& \geq \f 12 \log\lb 1+\f  {\al P} {Q  +1} \rb -3 \\
}
After $X_{IN}^N$ has been decoded, it is subtracted from $Y_I^N$ to estimate  $S_R^N$ through using the knowledge of $\theta^N$:
\ea{
\Yt_I^N=Y_I^N-X_{IN}^N =\sin(\theta^N) S_R^N+Z_I^N,
}
The codeword $X_{RC}^N$ is estimated from the vector
\eas{
\Yt_R^N
& = \sin(\theta^N) Y_R-\cos(\theta^N)\Yt_I^N \\
& = \sin(\theta^N) X_{RC}^N + \sin(\theta^N)Z_I^N-\cos(\theta^N)Z_R^N \\
& = \sin(\theta^N) X_{RC}^N + \Zt_R,
}
for
\ea{
\Zh_R= \sin(\theta^N)Z_I^N-\cos(\theta^N)Z_R^N  \sim \Ncal(0,1).
}
This corresponds to the equivalent channel without interference  whose capacity is
\ea{
R_{RC}
& =I(\Yt_R; X_{RC}| \theta),
}{\label{eq:rate aligment}}
for which, as in \eqref{eq:inequality sin capacity expression} we have
\ea{
R_{RC} \geq \f 12 \log \lb  \alb P + 1 \rb -1,
}
as already evaluated in \eqref{eq:inequality sin capacity expression} (for $Q$ instead of $\al P$).

\end{document}